# Conceptual Design of an HTS Dipole Insert Based on Bi2212 Rutherford Cable

**Alexander V Zlobin[1]\*, Igor Novitski[1] and Emanuela Barzi[1]**

[1] Fermi National Accelerator Laboratory, Pine and Kirk Rds. - Batavia 60510, IL, US; zlobin@fnal.gov, igor@fnal.gov, barzi@fnal.gov

\* Correspondence: zlobin@fnal.gov

**Abstract:** The U.S. Magnet Development Program (US-MDP) is aimed at developing high field accelerator magnets with magnetic fields beyond the limits of $Nb_3Sn$ technology. Recent progress with composite wires and Rutherford cables based on the first-generation high temperature superconductor $Bi_2Sr_2CaCu_2O_{8-x}$ (Bi2212) allows considering them for this purpose. However, Bi2212 wires and cables are sensitive to transverse stresses and strains, which are ~~quite~~ large in high-field accelerator magnets. This requires magnet designs with stress management concepts to manage azimuthal and radial strains in the coil windings and prevent degradation of the current carrying capability of Bi2212 conductor or even its permanent damage. This paper describes a novel stress management approach, which was developed at Fermilab for high-field large-aperture $Nb_3Sn$ accelerator magnets, and is now ~~is~~ being applied to high-field dipole inserts based on Bi2212 Rutherford cable. The insert conceptual design and main parameters, including the superconducting wire and cable, as well as the coil stress management structure, key technological steps and approaches, test configurations and their target parameters are presented and discussed.

**Keywords:** Bi2212; dipole coil; insert; Lorentz forces; mechanical structure; mirror structure

## 1. Introduction

Progress with round $Bi_2Sr_2CaCu_2O_{8-x}$ (Bi2212) composite wires, which can be used to produce Rutherford cables, makes them particularly suited for use in high-field accelerator magnets [1, 2]. This work started in the U.S. several years ago within the U.S. Very High Field Superconducting Magnet Collaboration (VHFSMC) [3] and is now being performed in the framework of the U.S. Magnet Development Program (US-MDP) [4]. Whereas the LBNL group, in collaboration with ASC-NHMFL-FCU, pursues coil inserts based on a Canted Cosine Theta (CCT) coil concept [5], Fermilab is focused on coil inserts based on the traditional cos-theta coil concept with stress management elements [6].

Several challenges are to be addressed on the way. The composite Bi2212 wire with Ag matrix is a soft and very delicate material which needs stringent empirical laws for Rutherford cabling to minimize the wire internal damage. Once the cable is formed and used to wind a coil, the Bi2212 coil requires a multistage heat treatment in Oxygen atmosphere at maximum temperatures close to 900°C to form the superconducting phase. Moreover, temperature homogeneity during the heat treatment has to meet strict gradient specifications. Similarly to $Nb_3Sn$, Bi2212 wires and cables are sensitive to stress and strain [7-9]. Although Bi-2212 is universally made with the Powder-in-Tube (PIT) technique, it appears that stress/strain behavior depends on the specific details of the technological processes used by each manufacturer. Pressure transverse to the cable face is one of the main stress components in accelerator magnets. Whereas very accurate measurements on tensile strain can be found in literature, more data on Bi2212 performance sensitivity to cable loading are needed. It is already certain, however, that stress management concepts will need to be applied to insert coils' designs when aiming at large magnetic fields.

The first step in developing a Bi2212 dipole insert, i.e. design studies, is complete. The Fermilab's insert is based on a 2-layer coil concept with small aperture. It uses Bi2212 Rutherford cable, which was available at LBNL. The coil fits into 60 mm aperture dipole coils available at Fermilab. The coil is wound inside a support structure. All parts for the coil support structure have been designed and will be fabricated by 3D sintering technology. As a first step, a practice coil using "dummy" $Nb_3Sn$ or Cu cable and 3D printed plastic parts will be wound, impregnated with epoxy and cut to examine the turn position in various parts of the coil. After winding, the real coil will be heat treated (reacted to form Bi2212 stoichiometry) at 50 bar in a furnace at NHMFL, and shipped back for impregnation with epoxy, instrumentation and cold testing at FNAL.

This paper describe the insert conceptual design and main parameters, including properties of the superconducting wire and cable, as well as the coil stress management structure, key technological steps and approaches, test configurations and their target parameters. In preparation for this and future inserts, cable development and characterization, including transport properties of extracted strands, cable at field, and cable under pressure will be performed. Some of these steps to be used for the first Bi2212 insert coil are reviewed in the paper.

## 2. Insert Coil Magnetic and Structural Design

### 2.1 2D Magnetic Design and Parameters

The design concept of the Bi2212 insert is based on a two-layer cos-theta coil with stress management concept which was developed at Fermilab for high-field large-aperture $Nb_3Sn$ accelerator magnets [6]. The coil uses a Rutherford cable with rectangular cross-section 7.8 mm wide and 1.44 mm thick, wrapped with 0.15 mm thick insulation [10].

This insert coil is designed to be tested inside 60-mm aperture $Nb_3Sn$ dipole coils developed at Fermilab [6, 11]. Thus, the coil outer diameter is limited to 59 mm, leaving 0.5 mm of radial space for the insulation between the Bi2212 and $Nb_3Sn$ coils. The coil inner surface is slightly elliptical with a minimal radius of 8.5 mm. There is ~2.4 mm of radial space outside of each coil layer for the mechanical support structure and inter-layer insulation. The coil consists of 15 turns, 5 turns in the inner layer and 10 turns in the outer layer. Each coil layer is split into 3 blocks, with the number of turns approximately following the cos-theta distribution. Each coil layer has a pole block, 2 inter-blocks and a small mid-plane spacers per quadrant, which ensure the radial turn position in the coil and minimize the low order geometrical harmonics.

The coil magnetic optimization was done using ROXIE [12] for a cylindrical iron yoke with inner diameter of 100 mm and constant iron magnetic permeability of 1000. The optimized coil cross-section is shown in Fig. 1. The coil geometrical parameters are summarized in Table 1.

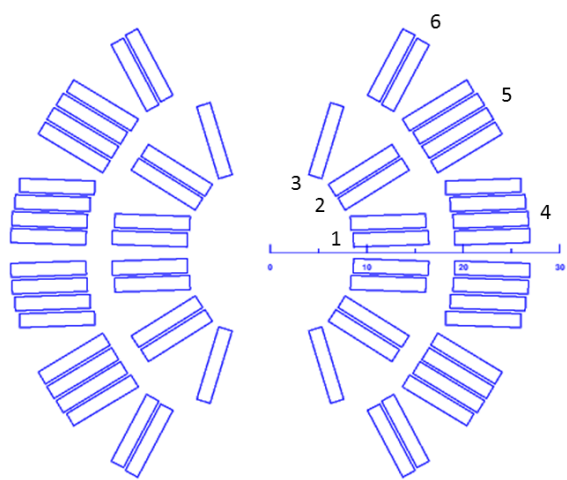

**Figure 1.** Optimized coil cross-section with block numbering in the 1st quadrant.

Table 1. Geometrical parameters of coil blocks.

| Block number | Number of cables in block | Block inner radius, mm | Block azimuthal angle, degree | Block inclination angle, degree |
|---|---|---|---|---|
| 1 | 2 | 8.50 | 3.0 | 2.0 |
| 2 | 2 | 8.79 | 28.6 | 32.0 |
| 3 | 1 | 8.83 | 53.0 | 73.0 |
| 4 | 4 | 19.0 | 2.0 | 2.0 |
| 5 | 4 | 19.0 | 25.0 | 31.0 |
| 6 | 2 | 19.0 | 48.0 | 62.0 |

Table 2: Design parameters of the Bi2212 dipole insert.

| Parameter | Value |
|---|---|
| Coil inner diameter, mm | 17.0 |
| Coil outer diameter, mm | 53.6 |
| Number of layers | 2 |
| Number of turns per half-coil (per layer) | 15 (5 IL+10 OL) |
| Maximum coil transfer function $TF=B_{max}/I$, T/kA | 0.57925 |
| Aperture transfer function $TF=B_o/I$, T/kA | 0.56839 |
| Maximum coil to aperture field ratio, $B_{max}/B_o$ | 1.019 |
| Coil inductance, mH/m | 0.3884 |
| Stored energy at coil current of 6 kA, kJ/m | 6.99 |

The calculated design parameters of the Bi2212 dipole insert are presented in Table 2. The length of the coil is limited by the available cable length of 15 m. It is estimated that the coil straight section will be ~300 mm long and the total length of coil winding will be ~450 mm, or approximately half of the Nb$_3$Sn coil length. Due to the small insert size, the coil cross-section area as well as the coil inductance, stored energy and Lorentz forces in this design are relatively small.

*2.4. Field Quality*

The induction of magnetic field $B(x,y)$ in the aperture of accelerator magnets is represented in terms of harmonic coefficients defined in a series expansion using the following complex functions:

$$B_y(x,y) + iB_x(x,y) = B_1 \sum_{n=1}^{\infty}(b_n + ia_n)\left(\frac{x+iy}{R_{ref}}\right)^{n-1},$$

where $B_x(x,y)$ and $B_y(x,y)$ are the horizontal and vertical field components, $b_n$ and $a_n$ are the normal and skew harmonic coefficients, $B_1$ is the main dipole component, and $R_{ref}$ is the reference radius. For this insert $R_{ref}$ = 5 mm, which corresponds to ~60% of the insert aperture radius.

The design low-order harmonics for the dipole central field, assuming an iron yoke permeability of 1000, are reported in Table 3. The cross-section of the insert coil, with the field uniformity diagram in the aperture and the field distribution in the coil cross-section at a current of 6 kA, is shown in Fig. 2. Whereas it was possible to reduce $b_3$, $b_9$ and $b_{11}$ harmonic coefficients below 1 unit, the other two low-order geometrical harmonics, $b_5$ and $b_7$, are relatively large, due to restrictions in turn positioning in this design. However, since they have opposite signs, the good quality field area, where $dB/B_1$ is smaller than 3 units, is round and close to 10 mm in diameter. It is shown as the dark-blue area in the aperture in Fig. 2.

**Table 3.** Geometrical field harmonics at $R_{ref}$= 5 mm.

| n | 3 | 5 | 7 | 9 | 11 |
|---|---|---|---|---|---|
| $b_n$, $10^{-4}$ | -0.76 | -9.6 | 3.43 | -0.23 | 0.03 |

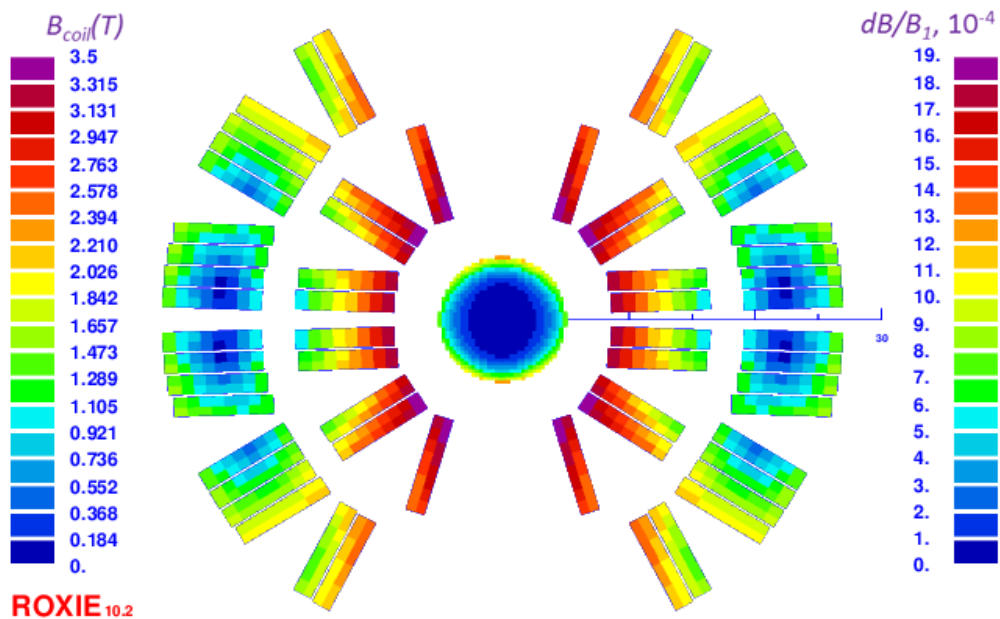

**Figure 2.** Cross-section of the insert coil with the field uniformity diagram in the aperture and the magnetic field distribution in the coil cross-section at a current of 6 kA.

*2.3 Coil Support Structure*

The Lorentz force values and directions in the coil blocks without any external field applied are shown in Fig. 3. At a coil current of 6 kA, the horizontal Lorentz force component $F_x$ = 97.5 kN/m/quadrant and the vertical one $F_y$ = -62.2 kN/m/quadrant. The Lorentz forces in the inner layer are practically horizontal, whereas in the outer layer there are both force components. Since the Lorentz force $F_L$ per unit length is a vector product of the current $I$ and the field induction $B$, or $F_L = I \times B$, the horizontal component of the Lorentz force in an external dipole field will proportionally increase, whereas the vertical one will not change.

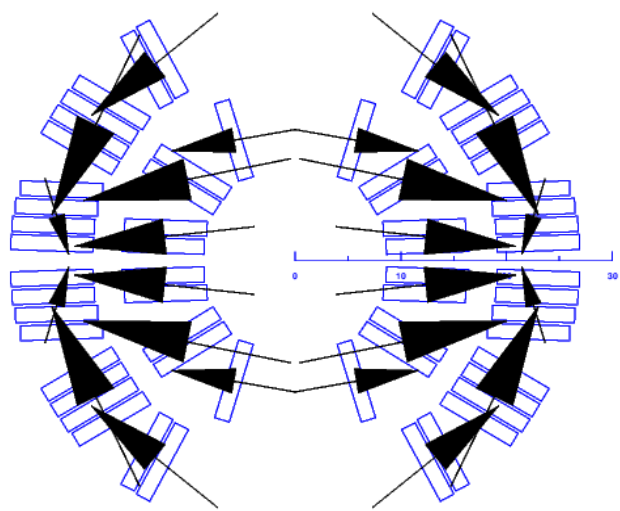

**Figure 3.** Lorentz forces in coil blocks at the coil current of 6 kA.

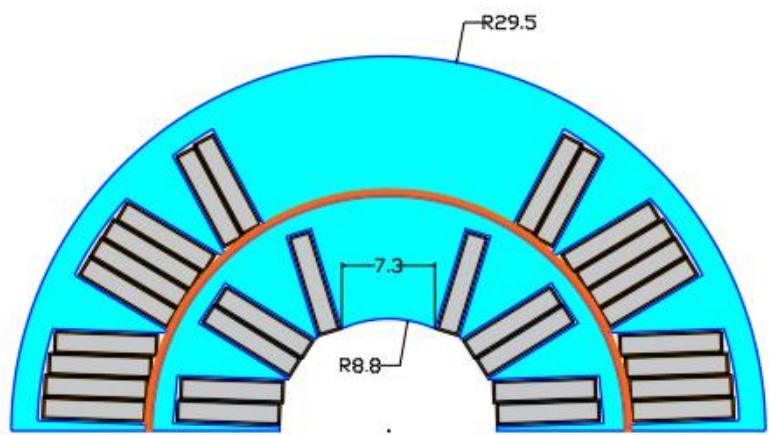

**Figure 4.** Cross-section of the coil inside the support structure.

Fig. 4 shows the Bi2212 half-coil inside the coil support structure. Each coil block is surrounded by an additional layer of insulation 0.15 mm thick. The coil support structure controls turn positioning during fabrication and operation, and protects the stress/strain sensitive Bi2212 cable from mechanical over compression during assembly and operation.

A 3D view of the coil turns in the non-lead end is shown in Fig. 5. Due to the very small aperture, the width of the inner-layer pole in this design is relatively small (~7.3 mm), which is quite challenging for the inner-layer pole-turn bending in the coil end areas. This important question will be addressed and optimized during practice coil winding.

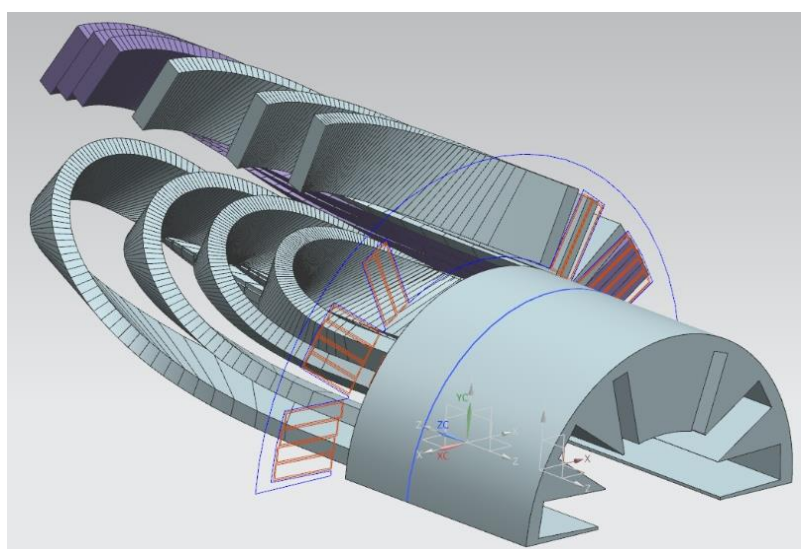

**Figure 5.** 3D view of coil turns in the coil non-lead end.

## 3. Materials and Technology

### 3.1. Wire and Cable Parameters and Technology

Cross-sections of the Bi2212 wire and the Rutherford cable are shown in Fig. 6. Their main parameters are summarized in Table 4. The Bi2212 wire was produced by Bruker OST LLC using precursor Bi powder provided by nGimat (now Engi-Mat). The cable was made and insulated at LBNL. The cable insulation consists of 0.15 mm thick mullite braided sleeve chemically compatible with the Bi2212 heat treatment process. The length of the insulated cable is ~15 m.

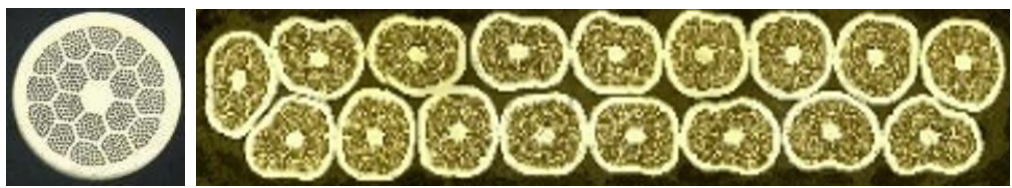

**Figure 6.** Bi2212 round wire (left) and 17-strand Rutherford cable (right) cross-sections (courtesy LBNL).

**Table 4.** Bi2212 cable and strand parameters.

| Parameter | Value |
| --- | --- |
| Cable ID | LBNL-1110 |
| Number of strands | 17 |
| Bare cable width, mm | 7.8 |
| Bare cable thickness, mm | 1.44 |
| Cable transposition pitch, mm | 58 |
| Billet ID | PMM180207-2 |
| Strand diameter before/after reaction, mm | 0.80/0.778 |
| Strand architecture | 55 x 18 |
| Strand fill factor, % | 23 |
| Strand twist pitch, mm | 25 |
| Strand $I_c(4.2K,5T)$ after NHMFL 50 bar OPHT, A | 460-640* |

* $I_c$ range within 10-degree heat treatment window, sample size - 15 samples

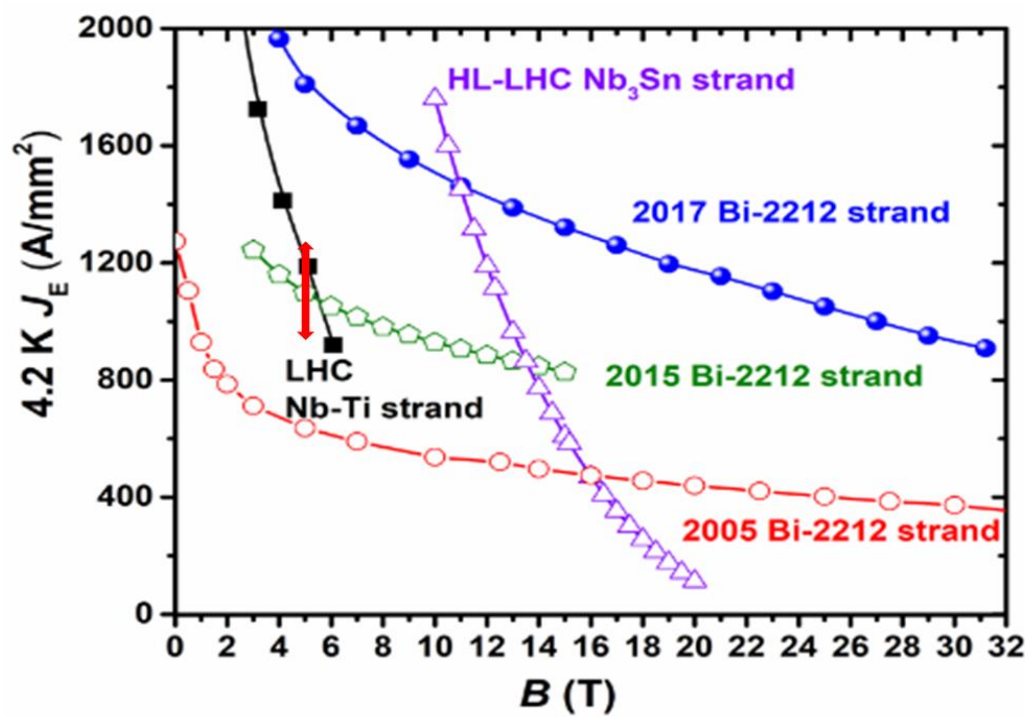

**Figure 7**. Improvement of the best of their time Bi2212 wire $J_e$ as compared with the LHC NbTi and HL-LHC Nb$_3$Sn wire specifications [10]. The 2005 curve represents best performance in the 2000s for Bruker-OST wires with Nexans powder; the 2015 curve shows similar wires' improvement in performance when undergoing the OPHT process; the 2017 curve is the best performance obtained by Bruker-OST with the new Engi-Mat powders. The range of $J_e$ for the wire used in the Rutherford cable provided by LBNL at 5 T is shown in the plot by the red double arrow, which is very close to the 2015 generation wire.

Bi2212 is a copper-oxide high temperature superconductor (HTS), which in addition to much higher critical temperature also have higher irreversibile magnetic field compared to low temperature superconductors. Bi2212 is the only copper-oxide material which can be easily melt processed and, thus, to be produced in a wide variety of shapes, including isotropic round multifilamentary wire. To achieve this so-called partial melt processing, a multistage heat treatment in Oxygen atmosphere at very uniform high temperatures up to 900°C is required. Bi2212 is made with the Powder-in-Tube (PIT) technique. Whereas the matrix embedding the filaments requires using pure Ag, the wire fabrication process typically uses an outer sheath made of Ag0.2%Mg alloy which is dispersion strengthened by oxidation of the Mg during wire heat treatment.

Significant progress was made in the development and industrialization of Bi2212 composite wires. Km-length quantities of Bi2212 round wires have been commercially produced since the 1990s. The critical current density $J_c$ and the engineering wire current density $J_e$ at 20 T and 4.2 K of the round wire increased from 1500 A/mm² and 400 A/mm² or less respectively in the 2000s, to the $J_e$ values shown in Fig. 7 today [10]. The range of the engineering wire current density for the wire used in the Rutherford cable provided by LBNL at 5 T is shown in the plot by the red double arrow, which is consistent with the parameters of the 2015 generation Bi2212 wire. The two main factors that led to the Bi2212 wire improvement are a) the removal of the 30% porosity in as-drawn Bi2212 wires by an overpressure processing heat treatment (OPHT), and b) the introduction of a new chemical powder technology by nGimat (now Engi-Mat), which produces highly homogenous Bi2212 precursor powders with very good composition control.

Fig. 8 gives an example of a typical heat treatment cycle, which is used for the partial melt processing at 1 bar of gas pressure. The gas is either pure Oxigen ($O_2$) or a mixture of Oxigen and Argon ($O_2$/Ar) with Oxigen partial pressure of 1 bar. The OPHT process uses 98% Ar and 2% $O_2$ with a total gas pressure at 50 bar, with the $O_2$ partial pressure still at 1 bar. Whereas the best wire performance is usually obtained by $O_2$ pre-annealing of the Ag0.2%Mg sheathed strand, the hardened sheath which results from this pre-anneal severely restricts the diameter around which the wire can be bent without cracking. To enable cabling of Bi2212 round stands, specific $O_2$ anneal processes have to be used. Then Bi2212 wires that pass a nominal bending test can be used to fabricate of Rutherford-type cables.

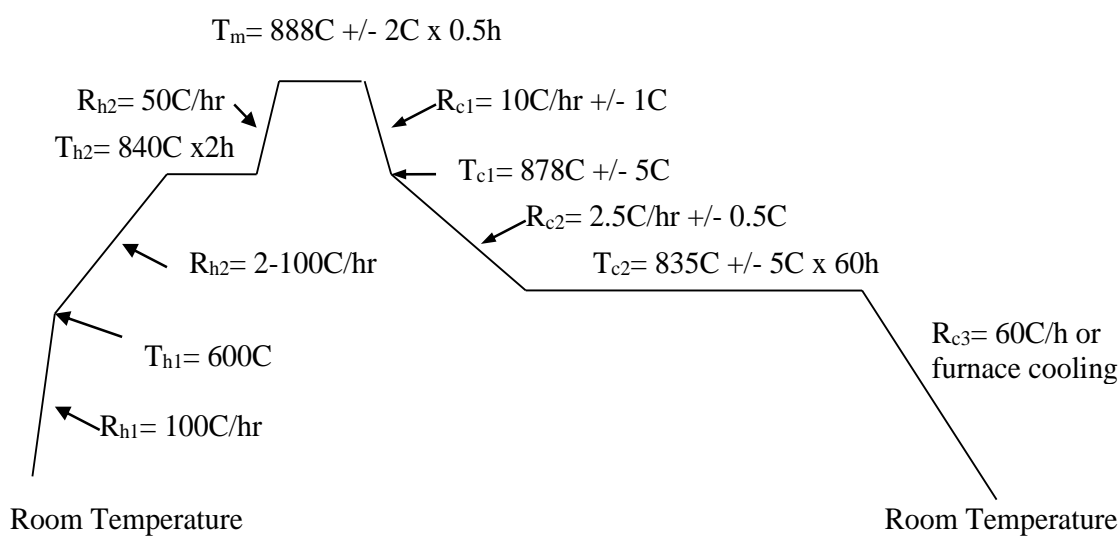

**Figure 8**. Bruker-OST optmized heat treatment for Bi2212 wire at 1 bar gas pressure.

Cable development is performed by designing and fabricating samples of different geometries using state-of-the-art wires, with the purpose of studying the effect of cable parameters and processing on their performance. This includes for instance the sensitivity of electrical properties (such as critical current $I_c$, residual resistivity ratio *RRR*) and internal structure (architecture, filament shape and spacing, sheath composition) to cable compaction, measurements of cable stability and AC losses, measurements of 3D cable expansion during the fabrication process and during reaction, effects of intermediate annealing when using a 2-pass fabrication process, etc. Rutherford cable finite element models [13] that evaluate for each considered cable geometry what is the plastic strain seen by the strands during fabrication, what are the most critical strand locations, and predicting local damage whenever the failure mechanisms of a specific strand technology are known, can also be used.

*3.2 Coil Structural Materials and Technology*

For the first insert coil, the coil materials and the current procedures developed at LBNL [10, 14] and at Fermilab will be used. The mandrel and the parts of the coil structure will be made of Aluminum-bronze-954 or Inconel-600, which is more expensive and challenging to machine, or some other appropriate material which can accept the high-temperature heat treatment in Oxigen atmosphere. The coil parts will be produced using precise 3D sintering technology. Every former part is pre-oxidized before the Bi2212 heat treatment, in order to create an $Al_2O_3$ oxidation layer that prevents the material from absorbing Oxygen from the environment during the coil heat treatment. The Bi-2212 Rutherford cable is insulated with mullite braided sleeve, i.e. ceramic braid 2 $Al_2O_3$:$SiO_2$. After winding and reaction the coil will be impregnated with epoxy. Whereas at LBNL, Mix-61 from NHMFL is used for this purpose, at Fermilab it will be done either using traditional CTD101K epoxy or some other tested impregnating materials under study at Fermilab, with their corresponding procedures. Coil materials and technologies are being carefully analysed and optimized for the coils of this series.

Unfortunately, the Bi2212 leaks that were observed at 1 bar reaction in Oxigen do not desist when performing the 50 bar heat treatment. Spots or discolorations form where Bi2212 liquid leaks through the encasing Ag alloy metal at high temperatures and reacts with surrounding materials, including the insulation [10]. Most leakages in small racetracks produced at LBNL occur at the Rutherford cable edges, and they degrade the $J_e$ locally. An example of leaks in Rutherford cables and their composition can be found in [15]. After reaction, the surface of all cables under study showed black spots embedded in the Ag coating as in Fig. 9, left and center. When tested at 4.2 K and self-fields of 0.1 to 0.3 T, an $I_c$ degradation of about 50% was found for all these cables. This current reduction on the cables was significantly and systematically larger than that of their extracted strands.

SEM/EDS analysis performed on the surface of a cable showed that the composition of this black material was very close to that of Bi-2212 (Spectrum 1 in Table of Fig. 9, at right). In a small crater at the edge of the sample, analysis showed several oxide phases, with specific shapes and morphologies, such as Bi-2212 (needle like grains), Bi-2201 (step like grains), (1,0) phase (spherical grains) and others. However, because no leaks were observed on the extracted strands, which performed well, the hypothesis was made that this problem is not as much related to the strand ability to withstand deformation as to the heat treatment of the cables itself.

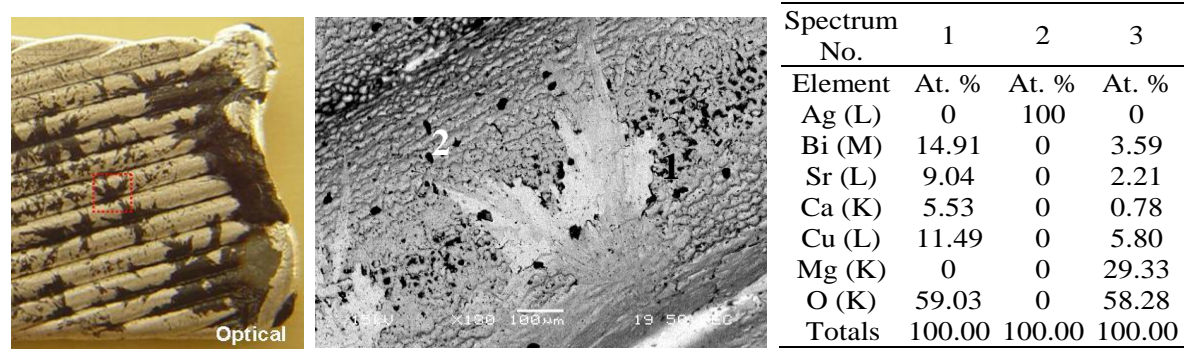

| Spectrum No. | 1 | 2 | 3 |
|---|---|---|---|
| Element | At. % | At. % | At. % |
| Ag (L) | 0 | 100 | 0 |
| Bi (M) | 14.91 | 0 | 3.59 |
| Sr (L) | 9.04 | 0 | 2.21 |
| Ca (K) | 5.53 | 0 | 0.78 |
| Cu (L) | 11.49 | 0 | 5.80 |
| Mg (K) | 0 | 0 | 29.33 |
| O (K) | 59.03 | 0 | 58.28 |
| Totals | 100.00 | 100.00 | 100.00 |

**Figure 9**. Bi-2212 Rutherford cable after reaction (left), back scatter image of circled black spot (center), and composition Table in marked locations (right) [15].

To reduce Bi2212 leaks, before insulating the Rutherford cable with the mullite sleeve, the cable is brushed with a thin layer of $TiO_2$-polymer slurry, i.e. $TiO_2$ powder mixed with ethanol. Once this is done, this coating is also applied to the insulated cable onto the mullite sleeve [10, 14]. This method does not eliminate leaks, but at least reduces their number.

*3.4 Effect of transverse pressure*

As previously shown, the main stress components on the conductor in the coil are transverse to the Rutherford cable axis. Transverse stress is the largest stress component in accelerator magnets, and can therefore damage brittle Bi2212 coils. To determine $I_c$ sensitivity to transverse pressure, electro-mechanical tests are typically performed on either cables or encased wires. Transverse pressure studies are made by applying pressure to impregnated Rutherford cable or wire samples and testing their transport current at several magnetic fields. Strain sensitivity increases with magnetic field.

There are two components of the critical current reduction: a reversible component, which is fully recovered when removing the load, and an irreversible component. The latter is permanent. The irreversible limit is defined as the pressure leading to a 95% recovery of the initial $I_c$ after unloading the sample. The $I_c$ degradation should strongly depends, as previously mentioned, on the Bi2212 wire technology.

The only data found in literature for transverse pressure testing of Bi2212 Rutherford cables are presented in [9]. The tested cable had 20 strands from a Showa billet, a Ni-Cr 80 core covered with two-layer wrap of MgO paper, and after reaction had expanded from 2.240 mm to 2.348 mm in thickness and from 8.89 mm to 8.94 mm in width. It was insulated with S-glass, placed in 304 stainless steel test tooling, and vacuum impregnated with CTD-101 epoxy. The cable sample was confined horizontally. The measurements were made at NHMFL, in a split-pair solenoid with a piston of 130 mm diameter in the loading device which pushed on the center of the beam, displacing it into the cable loading fixture. For a load of 100 MPa, mechanical modeling produced a displacement into the cable package of 25 μm, i.e. a strain on the Bi2212 cable of about 0.5%. Assuming an irreversible strain limit of 0.3%, in [9] it was recommended to limit the transverse pressure load to 60 MPa.

The transverse stress limit will be further verified by tests planned at Fermilab. The $I_c$ degradation should strongly depends, on the Bi2212 wire technology, but also on sample preparation and setup design. The former has an impact on possible stress concentrations; the latter determines the sample's actual stress–strain state. The Fermilab's Transverse Pressure Insert (TPI) measurement system is a device to test critical current sensitivity of impregnated superconducting cables to uniaxial (plane stress) transverse pressures up to about 200 MPa [13]. This device produces the effect of uni-axial and not multi-axial strain, since the experimental setup allows for the sample to expand laterally. This produces larger strain values on the cable sample than for instance on a laterally constrained one.

## 4. Coil insert testing

To test and optimize the Bi2212 coil design and technology, inserts will be tested independently and inside 60-mm aperture $Nb_3Sn$ dipole coils available at Fermilab.

The first single Bi2212 half-coil, described above, will be tested individually and in a background field of a 2-layer $Nb_3Sn$ coil that had been developed at Fermilab for the 11 T dipole for LHC [11]. The two coils will be placed inside the dipole mirror structure HFDM [16] as shown in Fig. 10.

Then the whole Bi2212 dipole insert will be tested inside a 4-layer 60-mm aperture dipole coil, which will feature stress management in the two outermost layers (see Fig. 11, left). This 6-layer hybrid coil will be placed inside a slightly modified 600 mm outer diameter structure (Fig. 11, right), which was developed and used at Fermilab for the MDPCT1 dipole [17]. The large-aperture coil with stress management has been developed [6] and is now being fabricated at Fermilab.

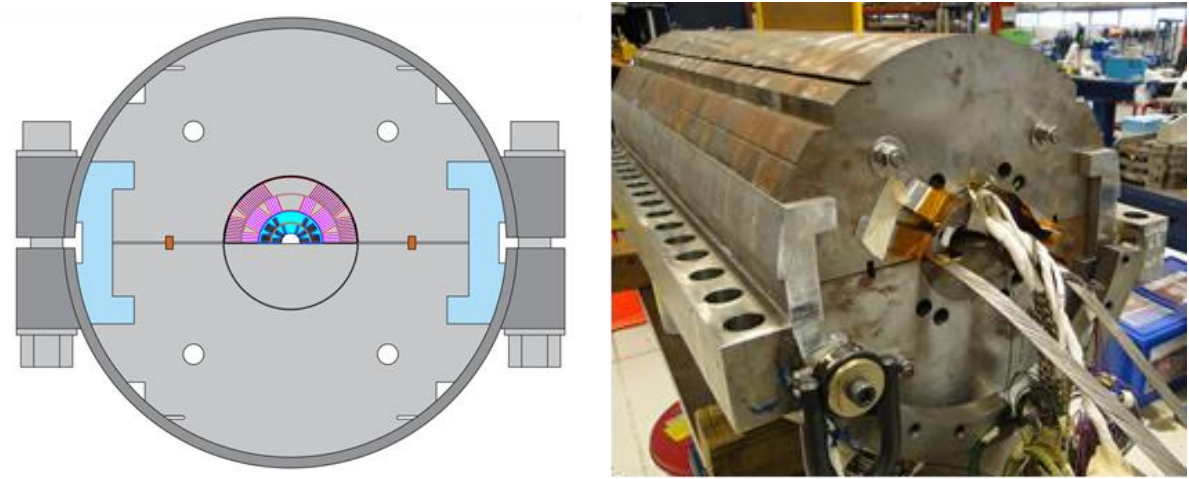

**Figure 10.** Bi2212 and 60-mm aperture 2-layer Nb$_3$Sn half-coils inside HFDM dipole mirror structure (left); and a picture of the HFDM mirror structure with 60-mm Nb$_3$Sn coil inside (right) [16].

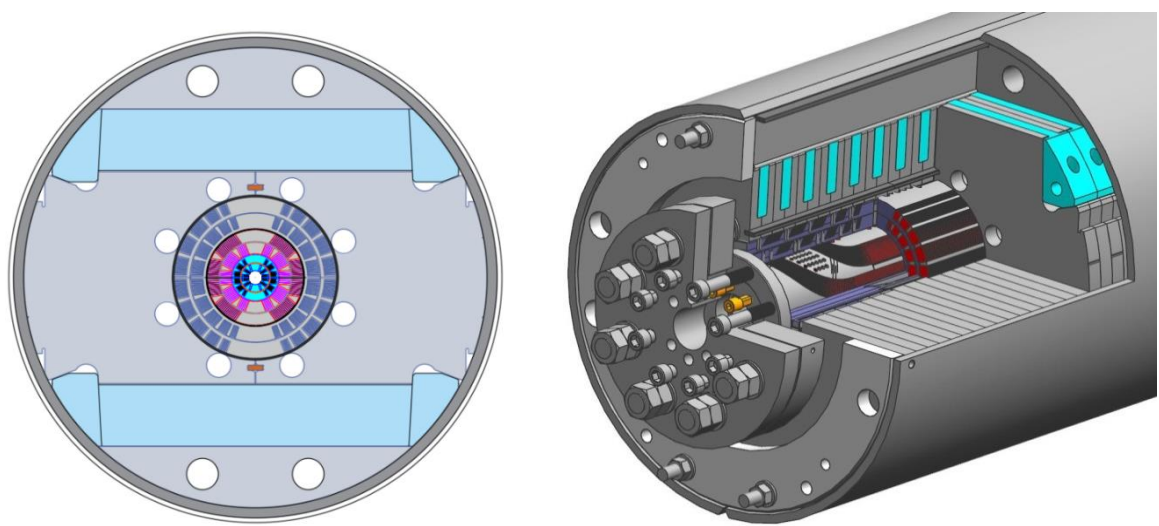

**Figure 11.** Cross-section (left) and 3D view (right) of 6-layer hybrid dipole coil inside MDPCT structure.

In both configurations, the Bi2212 coils will be powered individually and then connected in series with the Nb$_3$Sn coil. Since additional Lorentz forces from the Bi2212 insert coil are much lower than the Lorentz forces of the Nb$_3$Sn outsert coil, no modification of the structures shown in Figs. 10 and 11 is planned at the present time. Geometrical parameters of the hybrid test configurations discussed above are summarized in Table 5.

**Table 5.** Geometrical parameters of hybrid test configurations.

| Parameter | 4-layer dipole mirror | 6-layer dipole |
|---|---|---|
| Coil inner diameter, mm | 17.0 | 17.0 |
| Coil outer diameter, mm | 122.3 | 206.5 |
| Number of layers | 4 (2Bi2212+2Nb$_3$Sn) | 6 (2Bi2212+4Nb$_3$Sn) |
| Iron yoke outer diameter, mm | 400 | 587 |
| Maximum transverse size, mm | 545 | 613 |

The conductor limits of Bi2212 coils, made of two wire generations, inside the 4-layer and 6-layer hybrid configurations, were estimated using coil load lines and typical parameterizations for Bi2212 critical current $I_c(B)$, such as the following:

$$I_c(B) = \frac{A}{B^{0.2}}\left(1 - \frac{B}{B_{c2}}\right),$$

where $B_{c2}$ is 150 T and $A$ is 11.12 and 17.79 A·T$^{0.2}$ for Bi2212-2015 and Bi2212-2017 wires respectively. $I_c(B)$ curves of two Bi2212 wire generations and of the Nb$_3$Sn cable used in outserts, and maximum field load lines of the Bi2212 (1-4) and Nb$_3$Sn (5-6) coils are plotted in Fig. 12. The insert current and field limits for operation without and with external field from Nb$_3$Sn coils are defined by the intersection of the corresponding cable $I_c(B)$ curves with the insert load lines.

The maximum field in Bi2212 coils in various test configurations for two Bi2212 types of wires at various levels (from 30% to zero) of cable $I_c$ degradation is summarized in Table 6. The first coil made of Bi2212-2015 wire will be assembled and tested as a 4-layer hybrid mirror and powered independently and then in series with the Nb$_3$Sn coil outsert. For conductor degradation within 10-30%, the maximum field of the individually powered Bi2212 coil will vary within 3.3-4.1 T, and in the 4-layer hybrid mirror within 6.9-8.1 T. The whole Bi2212 dipole insert with the Bi2212-2015 wire tested as a 6-layer dipole, as seen from Table 5, could achieve a maximum field range of 9.9-12 T.

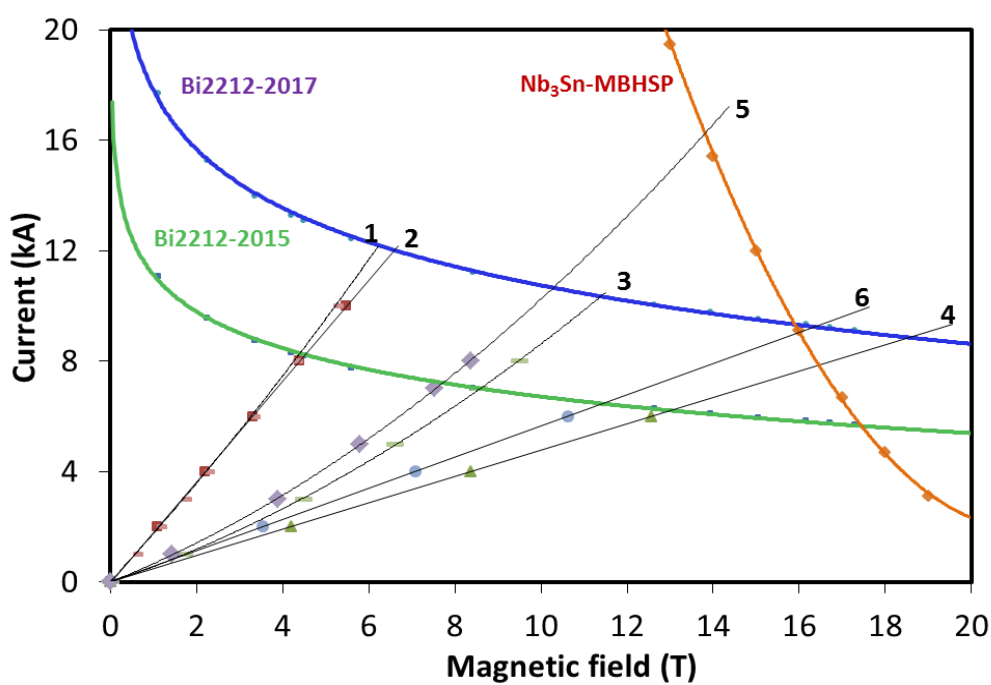

**Figure 12.** $I_c(B)$ curves and load lines of Bi2212 (1-4) and Nb$_3$Sn (5-6) coils in various test configurations: 1, 2 – single Bi2212 coil in dipole mirror and dipole; 3, 4 – Bi2212 coil in 4-layer hybrid mirror and 6-layer hybrid dipole; 5, 6 – Nb$_3$Sn coil in 4-layer mirror and 6-layer dipole.

**Table 6.** Maximum field in Bi2212 coil in various test configurations for two Bi2212 type wires at various levels of $I_c$ degradation in coils.

| Wire type | Test config. | Bi2212 coil | | | | Bi2212 + 2LNb$_3$Sn coils | | | | Bi2212 + 4LNb$_3$Sn coils | | | |
|---|---|---|---|---|---|---|---|---|---|---|---|---|---|
| | | $I_c(B)/I_{c0}(B)$ | | | | $I_c(B)/I_{c0}(B)$ | | | | $I_c(B)/I_{c0}(B)$ | | | |
| | | 1.0 | 0.9 | **0.8** | 0.7 | 1.0 | 0.9 | **0.8** | 0.7 | 1.0 | 0.9 | **0.8** | 0.7 |
| Bi2212-2015 | Mirror | 4.8 | 4.1 | **3.8** | 3.3 | 8.6 | 8.1 | **7.5** | 6.9 | - | - | **-** | - |
| | Dipole | - | - | **-** | - | - | - | **-** | - | 13.0 | 12.0 | **11.0** | 9.9 |
| Bi2212-2017 | Mirror | 6.2 | 5.8 | **5.3** | 4.8 | 11.4 | 10.7 | **10.0** | 9.2 | - | - | **-** | - |
| | Dipole | 6.6 | 6.1 | **5.5** | 5.0 | - | - | **-** | - | 18.6 | 17.2 | **15.7** | 14.2 |

It is expected that the next Bi2212 coils will use Bi2212-2017 wire. These coils could be assembled and tested in both the 4-layer hybrid mirror and in the 6-layer hybrid dipole configurations. The maximum field of the individually powered Bi2212 coil in the mirror will vary within 4.8-5.8 T. In the 4-layer hybrid mirror with the Bi2212 coil connected in series with the $Nb_3Sn$ coil, the maximum field will increase to 9.2-10.7 T for conductor degradation within 10-30%. The maximum field in the Bi2212 dipole insert, powered individually and in series with the 4-layer $Nb_3Sn$ outer coils as a 6-layer hybrid dipole, will reach 5.0-6.1 T and 14.2-17.2 T respectively for the same conductor degradation range.

As can be seen from Fig. 12, the critical current of the $Nb_3Sn$ coil in the 6-layer dipole is close but still higher than the critical current of Bi2212 coil. In the mirror configurations, the $Nb_3Sn$ coil has larger critical current than the Bi2212 coils. Notice also that the maximum field ranges of individually powered Bi2212 coils in the 4-layer mirror and in the 6-layer dipole configurations are very close since the load lines 1 and 2 in Fig. 12 are very close. This is not surprising since the field level produced by Bi2212 coils is relatively small and, thus, the iron saturation effect in the mirror is not large and the iron works as an ideal magnetic mirror.

## 5. Conclusions

The 2-layer dipole coil design, described above, allows extensive and cost-effective ways of developing and testing the technology of HTS inserts based on Bi2212 cable and the cos-theta coil geometry. With existing Bi2212 composite wires, there is the potential of reaching a maximum field in Bi2212 coils up to 16-17 T using a 6-layer hybrid dipole design. The maximum field in the coil bore will be ~2% lower, or within 15.7-16.6 T.

The presented insert design concept, as well as the basic technological solutions, will be studied experimentally on a series of short coils. The development of the Bi2212 coil engineering design is in progress. Tests of the first Bi2212 coil in dipole mirror configuration could start by end of 2021.

**Funding:** Work is funded by Fermi Research Alliance, LLC, under contract No. DE-AC02-07CH11359 with the U.S. Department of Energy and supported by the U.S. Magnet Development Program (US-MDP).

**Acknowledgments:** The authors would like to thank Dr. T. Shen and Dr. L. Garcia Fajardo from LBNL for providing Bi2212 cable parameters, and US-MDP Director Dr. S. Prestemon (LBNL) and Prof. D. Larbalestier (NHMFL/FSU) for their support of this work.